# Quality and Innovation with Blockchain Technology

Morgan C. Benton and Nicole M. Radziwill

## Abstract

In recent years, hype surrounding the proliferation of blockchain-based technology has been significant. Apart from the creation of bitcoin and other cryptocurrencies, it has been difficult to determine what practical utility might lie in the adoption of blockchain, mainly because there are so few in existence at present. Even so, interest in the technology has increased tremendously. This paper is a primer for software quality professionals. It briefly describes the history of blockchain technology, attempts to define and disambiguate terminology, fosters a general understanding of how blockchain works, and discusses how and why software quality professionals might want to invest time and energy in learning about, implementing, or using blockchain-based technologies in their own organizations -- or alternatively, improving the quality of blockchain technology itself.

## Introduction

In October 2008, a mysterious persona named Satoshi Nakamoto published a whitepaper called "Bitcoin: A Peer-to-Peer Electronic Cash System" on an internet mailing list. By January 2009, Nakamoto released version 0.1 of the Bitcoin software on Sourceforge. Although it was not backed by any government, existed as a purely digital product, and possessed no apparent intrinsic value, it began to be traded for goods and services of real value. The price of a bitcoin hovered under $10 USD for years, and then in early 2013 it underwent a sudden spike to over $100, then in late 2013, to over $1000, and then again in early 2017, it rapidly spiked again getting to over $5000 by September (coindesk.com, 2017). While it appears that the price of bitcoin is being driven up by a mix of financial speculation, and a rise in ransomware attacks where the attackers demand payment in bitcoin (Lee, 2017), the buzz around bitcoin has brought a lot of attention to the technology that serves as its foundation: the blockchain.

The hype surrounding blockchain technology has been intense over the last few years. Although the two technologies are very different, many people have confused blockchain with bitcoin, the cryptocurrency that made it famous. Furthermore, bitcoin's success has sparked the creation of nearly 1000 new cryptocurrencies (Wikipedia, 2017), and driven a craze for ICOs, or Initial Coin Offerings (Wilhelm, 2017), leading to the misconception that the only (or at least primary) application of blockchain technology is to the creation of cryptocurrency. Even critics of blockchain (e.g. Coppola, 2016) tend to emphasize the limitations of the technology from the perspective of the financial industry, rather than recognizing the broader implications of distributed ledger technology.

However, the blockchain is capable of supporting quite a bit more than cryptocurrency creation, and some of the newer platforms for blockchain development should be prompting forward-thinking software quality professionals to engage in innovation in this domain. Just over the past four years, research that includes the terms "blockchain," "quality," and "software quality" has become commonplace (see Figure 1). This paper will give a brief overview of the history of blockchain technology, describe how it works, and discuss examples of how it may influence and be impacted by professions.

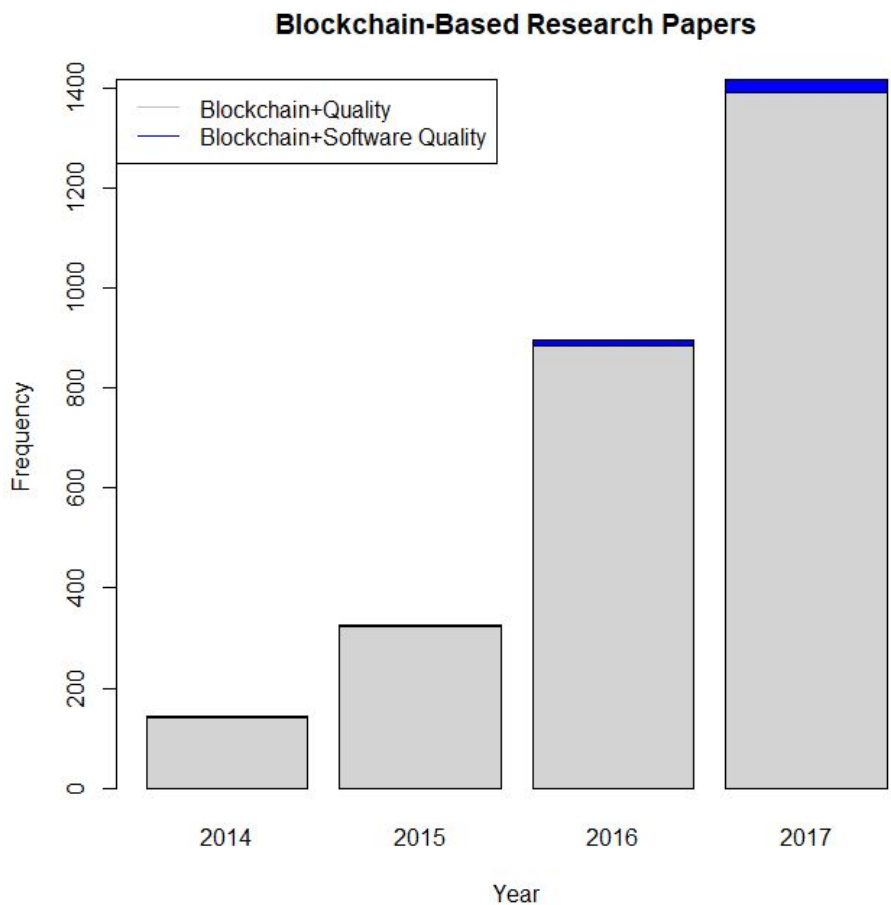

**Figure 1.** Frequency of papers in Google Scholar obtained by using the search terms (+"blockchain" +"quality" and -"software quality"), compared to (+"blockchain" +"software quality")

# History

As early as 1975, George Pake at Xerox PARC (and others) were already predicting the advent of a "paperless office" (Business Week, 1975). Nearly as predicted, by the early 1990s, it was clear that many or most documents moving forward would be stored digitally, and by 2003 one analysis showed that "the marginal cost of disk space for storing a document page is approaching 150 times less than the cost of the paper the document is printed on" (Hart & Liu, 2003, p97). Because it is easy to change digital files, research into how to maintain and ensure the validity and integrity of digital information was already well under way by the mid-1990s (e.g. Lynch, 1994). As the internet began to grow, people quickly understood that unless rock solid mechanisms of trust could be established, it would never realize its full potential, particularly in areas like commerce and governance.

The core concept of the "blockchain" was born when two researchers at Bellcore proposed "computationally practical procedures for digital[ly] time-stamping … documents so that it [would be] infeasible for a user either to back-date or forward-date [the] document" (Haber & Stornetta, 1991, p99). Shortly after that, they improved the technique so that multiple documents could be added simultaneously to a single block (Bayer, Haber, & Stornetta, 1993), and they also applied for a patent on the process (Haber & Stornetta, 1992). Far from obscure, their technique was noted and cited regularly in both academic and practitioner literature throughout the next two decades, including the *Handbook of Applied Cryptography* (Menezes, Van Oorschot, & Vanstone, 1996; Jansen & Karygiannis, 1998; De Roure, Jennings, & Shadbolt, 2001; Perrig, et al., 2005).

However, it was the introduction of bitcoin, a proposed peer-to-peer electronic cash equivalent (Nakamoto, 2009), that started the blockchain on its way to becoming a widely-known concept. It is important to understand that bitcoin is just one example of a product or service that is built upon blockchain technology. It was "the realization that the underlying technology that operated bitcoin could be separated from the currency and used for all kinds of other interorganizational cooperation" (Gupta, 2017), that catapulted the blockchain to prominence. While the story of bitcoin, which as of this writing has a market cap over $70 *billion*, is interesting in its own right, it is out of the scope of what the rest of this paper will address. The key thing to understand is that bitcoin and blockchain are *not* the same thing, although they are frequently discussed as if there were no difference.

Since the release of bitcoin in 2009, blockchain-based technologies have been on the rise. In addition to their use in cryptocurrencies (of which bitcoin is the most famous), there have been proposals and efforts to incorporate blockchain into a wide variety of products and services. The essential virtue of blockchain is the ability to automate mechanisms of **trust** without a central authority (like a central bank, government, or military), which mitigates risk, and enables all manner of efficiencies in human interaction whether in business or government contexts, whether formal or informal. "Smart contracts" are one of these technologies, promising to facilitate all manner of exchange of goods and services, not just financial ones. Alongside smart contracts, a great deal of effort is currently being placed in figuring out how to scale blockchain-based systems, and to implement them in a way that is much less computationally intensive than systems like bitcoin. Moving forward, there is hope that blockchain-based technologies will usher in a new wave of efficiency on a scale not seen since the internet boom of the last two decades (Gupta, 2017).

# Understanding the Technology

Given that hundreds of people are currently working on creating blockchain-based technologies, and there's a good chance these systems will eventually find their way into your industry, it is important to have at least a general understanding of how the blockchain works. This section will attempt to give you that understanding, and is geared towards someone who has a background in software development. You shouldn't need to be a cryptologist to understand this section, but conversely, it will not go into enough detail that you could implement a blockchain-based system yourself without a bit more study.

## Core Concept: An Immutable, Distributed, Digital Ledger

Simply put, a ledger is a record of transactions. Today, when people or organizations exchange goods, services, or currency for other goods, services, or currency, that transaction is frequently recorded in some sort of more or less durable medium, e.g. on paper, in a spreadsheet, or database. In those circumstances, the piece of paper, spreadsheet, or database constitutes the ledger. Examples include the amounts displayed or printed on your bank or credit card account statement, the deed of ownership for a piece of land recorded with a local government, or the amounts you write down in your checkbook (if you still use such a thing).

There exists an *enormous* global infrastructure of notaries, courts, and auditors whose primary purpose is to verify and validate that such ledgers are an accurate reflection of the world that they describe. Most of the time, the mere existence of this infrastructure is sufficient to ensure that all parties to a transaction honor the terms described in the ledger. However, humans spend a tremendous amount of time, energy, and resources preventing, monitoring, and working to resolve disputes between entities. Arguably, the primary function of government is to enforce the culture, laws, and rules described by these ledgers, aiming to preserve peace and order within and among our communities. As an example, if a person does not honor an agreement with a water or electricity provider by paying the bill (i.e. the ledger describing the transaction) on time, the service may be terminated. If the violation is egregious enough, police, lawyers, courts, jails, and prisons may get involved.

At its most basic, a **blockchain** is a shared, digital ledger that cannot be changed once a transaction has been recorded and verified. The algorithms used to carry out the verification and recording processes are implemented in software, and mathematically guarantee that once accepted, the details of the transaction described by the ledger cannot be altered by anyone, anywhere, without the application of more computing power than currently exists on the planet. All parties to the transaction, as well as a significant number of ostensibly neutral 3rd parties maintain a copy of the ledger (i.e. the blockchain), which means that it would be virtually impossible to alter *every* copy of the ledger globally to fake or cheat on a transaction.

It is critical to note a couple of key caveats. First, the mere existence of a transaction in a blockchain does not necessarily guarantee that it is a true representation of the interaction between two entities--people and organizations are still susceptible to being fooled, careless, or misled into entering an otherwise legitimate transaction. Second, even though the blockchain is designed to be incontrovertible proof of an agreement between two parties, there is no guarantee of retribution, remuneration, sanction, punishment, or any other consequence should the societal mechanisms of enforcement fail to operate for some reason, such as corruption, apathy, or simply being overwhelmed -- i.e. even though I may be able to prove that you owe me money, that doesn't necessarily mean that someone will force you to pay.

That being said, with proper design and implementation, the probability that these shortcomings will cause problems can be greatly minimized. For example, current implementations of software for facilitating bitcoin transactions (i.e. bitcoin clients) require parties to authenticate themselves using 2-factor authentication, the use of 3rd party authentication systems (like Google Authenticator), as well as IP address verification before any transfer of bitcoin from one wallet to another can be executed. There is also a 48-hour waiting period that is accompanied by email verification. The goal of all of this being to reduce as much as possible the likelihood that someone will be a victim of theft or other fraudulent transaction. A great deal of current work on blockchain-based technologies resides in minimizing the likelihood of fraud or cheating. Likewise, another area of focus for development involves automating the means of enforcing or reverting a transaction should one of the parties fail to live up to their side of the bargain.

Finally, it is important to be clear about what is meant by the word "distributed." In the context of blockchain technology, distributed generally means that there is no central repository or canonical version of the ledger. Every member of the network possesses an equally legitimate version of that ledger. The very fact that there is no central authority is a big part of what makes blockchain attractive--in a very real sense, a core value of autonomy is baked into every blockchain application. As will be discussed later, this is both extremely attractive and simultaneously very scary to governments and other power-brokers in the world today.

## Blocks: The Units of Content in a Ledger

As the name implies, a blockchain is made up of a chain of "blocks." The content of these blocks is determined by the developers of the blockchain client software. For example, with bitcoin, the content of each block is a list of transactions of bitcoin moving between digital wallets. However, the content of a block could be *anything* that could be represented digitally, including photographs, video, audio, or anything else that can be digitized. In the case of Ethereum, the content of a block is actually a piece of executable software that executes a contract (Lewis, 2016). As described above, it is up to the software developers to come up with a way to prevent content from being encoded in a block that is not an accurate representation of an actual transaction between entities. All of that said, what is the mechanism that ensures that the contents of a block become immutable and verifiable?

Blocks are typically encrypted using public key cryptography. Public key cryptography has been around since the 1970s, and is the foundation for most of the security mechanisms employed by modern computing systems, including SSL, the technology that protects the privacy of a great deal of the data transferred across the internet. Users of a blockchain technology typically will create a public-private key pair. The public key is shared with other users on the network and can be used to encrypt information intended only for the owner of the public key. Once received, the owner of the public key must use both the public and private keys to decrypt the information.

In the context of blockchain technology, the user's keys are used to digitally sign a contract or transaction. The signature is based upon not just their own keys, but also upon the digital signature of the most recent block in the chain. That way, once a file or contract has been "signed," all of the other clients on the network are able to verify that the content may be added to a new block. Whenever a new block is created, all of the transactions that have occurred since the creation of the most recent block are bundled together and recorded as a new block. It then should be possible for anyone using the application to determine the signatures of the entities involved in a transaction, as well as to verify the contents of that transaction.

The order of blocks is significant. As the concept of a "chain" implies, all of the blocks are linked to one another in a fixed, unchangeable order, that is determined by the time at which the block is created. The first block in the chain is referred to as the "genesis" block, and is generated with a "seed" key. The digital signature of the genesis block is combined with all of the transactions that have occurred since it was created and used to generate the second block. Likewise, the signature of the second block is combined with subsequent transactions to create the third block, and so on.

The entire series of blocks constitutes the blockchain, and is distributed to all of the members of the network. As more blocks are added to the chain, each client must have a mechanism to receive, verify, and record those blocks. As described above, each member of the network is independently responsible for, and participates in the verification of every block that is added to the chain. If any entity attempts to modify earlier transactions in the blockchain, it will be immediately detectable by all of the other nodes on the network. As might be guessed, as the blockchain grows in length, and as the network grows in members, it requires more and more storage space, and more and more computing power to compute and store the blockchain. Figuring out ways to make blockchains more scalable is currently a hot topic of research.

All of this begs the question, if the blocks encode transactions, who creates the new blocks and why would they do so? This is one of the true genius ideas of blockchain technology. Members of a blockchain network are rewarded for contributing resources toward the computation of subsequent blocks in the chain. For the bitcoin network, the resource that was contributed was computing power, the reward for contributing was bitcoin, and the act of contributing computing power to the network was referred to as "mining." Each time a computer successfully "mined" a new block for the bitcoin blockchain, the miner was rewarded with a certain number of bitcoin (currently 12.5 bitcoin, which has an approximate value of $50,000 USD). The evidence of the mining was referred to as "proof of work," since it was not possible to discover the next block in the chain without expending a certain amount of computing power. The complexity of the work was continually adjusted so that it would always be about ten minutes between when one bitcoin block was mined and the next.

Since mining blocks requires the investment of resources, and since it is governed by a set of rules (which are set by the creator of the blockchain), people who would abuse the system are strongly discouraged from participating. In other words, the rewards for being a "good" miner who plays by the rules, greatly outweigh the cost involved with cheating the system. These forces perpetuate participation, and ensure the continuation of the blockchain.

There have been a number of problems with the particular implementation of the bitcoin blockchain. For example, the use of computing power as "proof of work" means the cost of mining new bitcoin blocks is tied to the amount of electricity needed to power the servers used to mine those blocks, which is tied to some amount of fossil fuels used to generate that electricity. As such, mining bitcoin encourages the burning of fossil fuel for no other reason than to mine bitcoin, which is not considered environmentally sustainable. A lot of current research and development revolves around coming up with alternative ways to mine blocks. "Proof-of-stake" is one of the more promising ones that uses an entity's interest in the outcome of transactions as the guard against foul play.

## The Network

Blockchains would not exist without computer networks, such as the internet. It is by way of networks that all the users of a particular blockchain technology are linked, and it is via networks that the ledger is distributed and maintained. While this has been implied in the previous sections, it bears a bit of examination on its own. While in the majority of instances the network that gets used for transactions is the internet, this is not a requirement, and there have been suggestions that alternative networks--the "dark web," VPNs, etc.--might be used. It is also conceivable that organizations might have an internally used blockchain that only exists and has meaning within their organization.

In the case of bitcoin, the use of the internet became a threat to the viability of the blockchain at one point. Once adoption of bitcoin had reached a critical mass, it became clear that there was advantage to amassing enough specialized computational power to make it virtually impossible for any other competitors to mine bitcoin successfully. As it happened, the first people to realize this who had the resources to build a massive server farm solely for mining bitcoin resided in China. There was fear that if the Chinese government, who maintains a massive firewall around the country, decided to do so, they might interfere with the ability for all of the nodes on the bitcoin network to communicate, and hence stall the growth of the blockchain (Antonopoulos, 2014).

## Potential Uses for Blockchain

By this time, it should be clear that blockchain technology can be applied to many more problems than just cryptocurrency. In their step-by-step guide, BlockGeeks (2017) outline several potential uses, including:

- **Smart Contracts**: Programs that execute only when specific conditions have been met.
- **A True "Sharing" Economy:** Cutting out intermediaries like Uber completely, and letting individuals exchange things of value directly, without overhead or brokers.
- **Crowdfunding:** Going beyond the model of Kickstarter and IndieGoGo to allow funders to participate in the management of projects they back with voting rights earned by their contribution.
- **Governance:** There is huge potential to use blockchain to manage online voting, for elections as well as smaller matters, and create true participatory democracy -- giving everyone a direct say in the use of shared resources.
- **Supply Chain Auditing**: Give supply chain partners and consumers a way to verify the origin of products and component materials, e.g. that "green" products are actually sourced from environmentally conscientious suppliers.
- **Personal Data Management**: Today, platforms like Facebook and Twitter profit by selling users' attention to advertisers. Blockchain might enable micropayments to accrue to users in exchange for access to their attention and other personal information, while providing enhanced layers of personal control.
- **AML and KYC**: Anti-Money Laundering (AML) and Know Your Customer (KYC) rules cost financial institutions, and in turn customers, huge sums. Blockchain could automate currently labor intensive work.

There are many more applications for blockchain than just those that are listed here. The next section will focus specifically on ways that blockchain intersects the software quality professions: 1) how to achieve quality assurance in a blockchain, 2) how to apply a blockchain to increase software quality, and 3) how to use blockchain technology to support continuous improvement.

# Quality Issues and Blockchain

The blockchain concept emerged from the concepts of *distributed systems* and *peer-to-peer networks.* In distributed systems, the number of participants and their identities are not only well known, but (to some degree) controllable. Peer-to-peer systems deviated from this model by enabling any participant to join the network, facilitating capabilities like redundant file storage and distributed access (e.g. Napster, BitTorrent). These "permissionless" systems assume that most of the participants are honest. Though robust to connectivity issues and other failures, they are particularly prone to cyberattacks in which one threat actors spawns many participants in the network. In addition, there is no good way to ensure that participants receive the resources they are expecting, nor is there a mechanism to reduce or eliminate variability: two participants making the same request may receive entirely different resources, or may receive the same resource when this is not permitted (Pass et al., 2017).

Blockchain solves these problems by inserting puzzles to be solved into the process (the "proof of work" concept). Still, there are many quality-related issues surrounding this technology. The following sections explore two themes: quality assurance of the blockchain *itself*, and quality that *results from* implementation of a blockchain.

## Quality of the Blockchain

Understanding the required quality attributes of blockchain implementations is still a topic of active research; there are very few researchers focusing on these issues. As a "public digital and distributed database solution providing decentralized management of transaction data" (p8:1), blockchain uniquely operates on a peer-to-peer network where no node has greater authority than any of the other nodes (Koteska et al., 2017). These authors have provided the only comprehensive examination of quality in blockchain to date; they recommend continuous testing, and conclude that blockchain implementations "need to be improved in terms of scalability, latency, throughput, cost-effectiveness, authentication, privacy, [and] security" (p8:7).

Other researchers have emphasized specific quality attributes, especially those that relate to security and fairness. Pass et al. (2017), for example, found that network delays can be particularly pernicious: rogue actors can potentially create denial of service attacks that generate gaps where they can solve puzzles while the honest actors wait. To solve this "fairness" problem, and ensure that all participants have equal opportunity to contribute to the puzzle-solving process, Pass & Shi (2017) have developed a variant of blockchain called "FruitChain" which specifically *decreases the variance* associated with assignment of rewards.

Blockchain deployment interfaces that currently exist do not have built-in fault tolerance for either connectivity issues or execution errors. In addition, insecure clients have directly resulted in losses for blockchain-based cryptocurrency systems, an issue that is compounded by the fact that developers know about this problem and have chosen not to document it (Walker et al., 2017).

Janze (2017) examines quality issues in decentralized information systems in general, motivated by the promise of increased use of blockchain. He finds ten factors that influence the quality of a strongly decentralized information system: societal norms, economic boundaries, intention to contribute, intention to use, objective quality, perceived quality, level of contribution, level of usage, intellectual net benefit, and economic net benefit. This model goes well beyond the observations of Koteska et al. (2017) who emphasized the perspective of a single blockchain implementation only.

## Using Blockchain to Improve Data Management and Validation

Blockchain is also presented as a means to solve data quality issues at many scales. A very simple example would be using blockchain to validate that the person who posted a message on a social media system (like Twitter) is who you think it is (Snow et al., 2014). All this would require is for the message to be signed by a private key, and everyone would benefit by the absence of "sybils" or fake (bot) accounts which can be used to spread misinformation or launch cyberattacks. If carefully architected, blockchain technology could also potentially be used to stop the spread of "fake news" online--although this is a challenging topic because it involves the behavior and cognitive biases of people in addition to technology.

Protecting critical data is one area where blockchain could be particularly strong, for example in cases where criminals or other threat actors gain access to databases with sensitive information. In a blockchain-based system, the nature of the threat is completely different. Hash values containing sensitive data are stored in a blockchain that *itself* is distributed over multiple machines--hackers would only be able to retrieve bits and pieces of the personal information, and would not have the capability to reconstruct it to gain leverage (Cheng et al., 2017).

The data management benefits are also likely to be critical for a secure Internet of Things (IoT). Giannetsos & Michalas (2016) point out that there are many critical security issues associated with IoT, especially when those components gather information about their users or the environment in which they are deployed. Sharing information, even when the information itself is not sensitive, may be associated with disclosing metadata like geographical location which could (in extreme cases) endanger the user. Furthermore, sensor data gathered by IoT users (called "participatory sensing") must be transmitted, shared with machine learning algorithms, and stored in such a way that the storage provider is not exposed to inordinate liability. This sensitive ecosystem of identity management, data sharing, and secure data storage could potentially be more easily navigated with blockchain-based implementations.

## Using Blockchain for Continuous Improvement

The traceability characteristics of the blockchain are very desirable for applications in supply chain management, where provenance and supplier quality assurance are paramount. Supply chain visibility will be particularly helpful for managing components like electronic chips (Daskalos, 2015). Alone, these components are typically not expensive, but if they are faulty or contain malware, they can cause serious downstream failures in products and even critical infrastructure (e.g. energy production, water/wastewater management).
Korpela & Hallikas (2017), studying attitudes of managers surrounding the concept of "digital supply chain," explored how large, heterogeneous, distributed data repositories could be organized to generate value--specifically by tighter online integration between trading partners. What they describe is identical to the value added by Electronic Data Interchange (EDI) systems in the 1980's and 1990's. They concluded that the value of blockchain may be inhibited by other issues that are not new: (lack of consistent) standards, (inconsistent) timestamping of transactions, monitoring and tracking of information flows, and secure end-to-end information transfer.

Beyond supply chain management, auditing, retail operations, and business process management may also benefit from blockchain-based technology. The job of auditors may be reduced or eliminated; if the system checks the viability and accuracy of all transactions in real-time, there should be no need to go back and revisit what transactions occurred. Chakrabati and Baudhuri (2017) note that innovative customer loyalty programs may emerge, more effective customer profiling may be enabled, and better validation of product authenticity may change the nature of business models. The processes within operations could also be enhanced: Weber et al. (2015) explored the benefits of "collaborative process execution" using blockchain. In a simulation, they created 500 smart contracts and 8000 transactions, and found that monitoring and coordination of business processes was feasible without any central authority. Conceivably, decentralized continuous improvement would also be a possibility.

These systems will make it more difficult and more expensive to cheat, while potentially providing specific and immediate incentives for continuous improvement. The psychological landscape of continuous improvement may, as a result, shift.

## Using Blockchain to Improve Software Quality

Because of its potential utility as a mechanism for managing a real-time (and fully automated) audit process, blockchain technology may be useful as the basis for next-generation regression testing systems. In addition, Porru et al. (2017) believe that new modeling techniques (e.g. a Unified Modeling Language for blockchain implementations), specific design notations or reference architectures, and blockchain-specific software development methodologies (e.g. an updated Cleanroom technique) may be required. Better methods to evaluate the structure and details of smart contracts will also be required. Idelberger et al. (2016) have recommended declarative or logic-based languages to accomplish this task. In all cases, testing would require simulating the entire blockchain to test against.

It is also possible that disruptive innovation will occur related to the *nature* of the distributed testing process as well. Walker et al. (2017), for example, explored the requirements associated with tools and techniques to enhance management development, deployment, execution, and testing of blockchain applications for the Internet of Things by considering a prototype for next-generation energy markets. In the envisioned energy ecosystem, which has also been described in the 2011 and 2015 ASQ Future of Quality Study (ASQ, 2015), consumers can produce energy in addition to consuming it, and contribute excess energy back into the market for others to purchase. By analyzing a software system to support this potential new transaction model, they found that it was possible to create "repeatable testing networks" that demonstrated scalability.

Alternatively, a blockchain-based system might be used to support the development of an entirely new project: if an organization develops specifications and builds unit tests, developers from anywhere in the world could provide portions of the finished code as their "proof of work," and the identity management and transaction processing capabilities of the blockchain could be used to find and compensate them for their services, no matter how small or incremental the contribution. Such an approach could potentially be a breakthrough innovation for managing software development, catalyzing diversity in the workforce by decoupling the full-time work contract from the finished product, and eliminating bureaucratic requirements that accompany even part-time contract labor.

## Conclusion

Blockchain technology is relevant not only for financial tracking and management, but potentially also for transactions that require personal identification, peer review, democratic decision-making or consensus, or solid provenance and audit trails. It may reduce information asymmetries between people, increase transparency, and ultimately help build trust between people and autonomous systems. Although the technology is more accessible now than it was a few years ago, more research and exploratory development is needed to translate the potential of blockchain into real value across many industries. It is not yet ready for off-the-shelf implementations, and a scarcity of understanding may be inhibiting its adoption. This article aimed to provide a solid conceptual introduction to the nature and function of blockchain technology that will prepare readers to implement and continuously improve it as needed to meet organizational objectives.